\documentclass[twocolumn,superscriptaddress,aps,nofootinbib,
showpacs,preprintnumbers,lengthcheck]{revtex4-1}
\usepackage[english]{babel}
\usepackage{amssymb}
\usepackage{amsmath}
\usepackage{enumerate}
\usepackage{graphicx}
\usepackage{dcolumn}
\usepackage{bm}
\usepackage{hyperref}
\usepackage{makeidx}
\usepackage{placeins} 
\usepackage[usenames,dvipsnames]{color}

\makeindex

\begin{document}

\title{Static and dynamic friction of hierarchical surfaces}

\author{Gianluca Costagliola}
\affiliation{Department of Physics, University of Torino, Via Pietro Giuria 1, 10125, Torino, 
Italy.}
\author{Federico Bosia} 
\affiliation{Department of Physics, University of Torino, Via Pietro Giuria 1, 10125, Torino, 
Italy.}
\author{Nicola M. Pugno}
\email[Corresponding author:]{nicola.pugno@unitn.it}
\affiliation{
Laboratory of Bio-Inspired \& Graphene Nanomechanics, Department of Civil, Environmental \\and 
Mechanical Engineering, University of Trento, Via Mesiano, 77, 38123 Trento, Italy}
\affiliation{Center for Materials and Microsystems, Fondazione Bruno Kessler, Via Sommarive 
18, 38123 Povo, Trento, Italy}
\affiliation{School of Engineering and Materials Science, Queen Mary University of London, 
Mile End Road, London E1 4NS, UK}

\date{\today}

\begin{abstract}
Hierarchical structures are very common in Nature, but only recently have they been 
systematically studied in materials physics, in order to understand the specific effects they can 
have on the mechanical properties of various systems. Structural hierarchy provides a way to tune 
and optimize macroscopic mechanical properties starting from simple base constituents, and new 
materials are nowadays designed exploiting this possibility. This can be also true in the 
field of tribology. In this paper, we study the effect of hierarchical patterned surfaces on the 
static and dynamic friction coefficients of an elastic material. Our results are obtained by means 
of numerical simulations using a 1-D spring-block model, which has previously been used to 
investigate various aspects of friction. Despite the simplicity of the model, we highlight some 
possible mechanisms that explain how hierarchical structures can significantly modify the friction 
coefficients of a material, providing a means to achieve tunability.
\end{abstract}

\maketitle

\section{Introduction}\label{intro}

The constituent laws of friction are well known in the context of classical mechanics, 
with Amontons-Coulomb's (AC) law, which states that the static friction force is proportional to 
the applied normal load and independent of the apparent contact surface, and that the kinetic 
friction is independent of the sliding velocity \cite{persson}. This law has proved to be 
correct in many applications. However, thanks to advances in technologies, with the possibility to
perform high precision measurements and to design micro-structured interfaces, its validity range 
was tested and some violations were observed in experiments, e.g. \cite{exp1}\cite{exp2}. 
Indeed, despite the apparent simplicity of the macroscopic laws, it is not easy to identify the 
origin of friction in terms of elementary forces and to identify which microscopic degrees of 
freedom are involved.

For these reasons, in recent years many models have been proposed \cite{rev}, additionally incorporating 
the concepts of elasticity of materials, in order to explain the macroscopic friction properties 
observed in experiments and to link them to the forces acting on the elementary components of the 
system. Although many results have been achieved, it turns out that there is no universal model 
suitable for all considered different materials and length scales. The reason is that the 
macroscopic behaviour, captured in first approximation by the AC friction law, 
is the result of many microscopic interactions acting at different scales.

As pointed out in the series of works by Nosonovsky and Bhushan \cite{nos1}\cite{nos2}, 
friction is intrinsically a multiscale problem: the dominating effects change through the 
different length scales, and they span from molecular adhesion forces to surface roughness contact 
forces. Hence, there are many possible theoretical and numerical approaches, depending on the 
system and the length scales involved (see ref. \cite{rev} for an exhaustive overview). 

The situation is much more complicated if the surfaces are designed with patterned
or hierarchical architectures, as occurs in many examples in Nature: the
hierarchical structure of the gecko paw has attracted much interest 
\cite{gecko_nature}-\cite{gecko_gorb}, and research has focused on manufacturing artificial 
materials reproducing its peculiar properties of adhesion and friction. In general, the purpose of 
research in bio-inspired materials is to improve the overall properties (e.g. mechanical) by 
mimicking Nature and exploiting mainly structural arrangements rather than specific chemical or 
physical properties. In this context, nano and biotribology is an active research field, both 
experimental and theoretical \cite{nanotrib1}-\cite{snake}.

Since hierarchical structures in Nature present such peculiar properties, it is 
also interesting to investigate their role in the context of tribology, trying to understand, for 
example, how structured surfaces influence the friction coefficients. This can be done by 
means of numerical simulations based on ad-hoc simplified models, from which useful information can 
be retrieved in order to understand the general phenomenology. From a theoretic and 
numerical point of view, much remains to be done. For this reason, we propose a simple model, i.e. 
the spring-block model in one dimension, in order to explore how macroscopic friction properties 
depend on a complex surface geometry.

The paper is organized as follows: in section \ref{sec_mod}, we present the model in details and 
we discuss results for non structured surfaces, that are useful to understand the basic 
behaviour of the system. In section \ref{sec_patt}, we present results for various types of 
patterned surfaces. In section \ref{conc}, we discuss them and provide the conclusions and 
future developments of this work.

\newpage

\section{The Model}\label{sec_mod} 

\begin{figure}[h!]
\begin{center}
\includegraphics[scale=0.19]{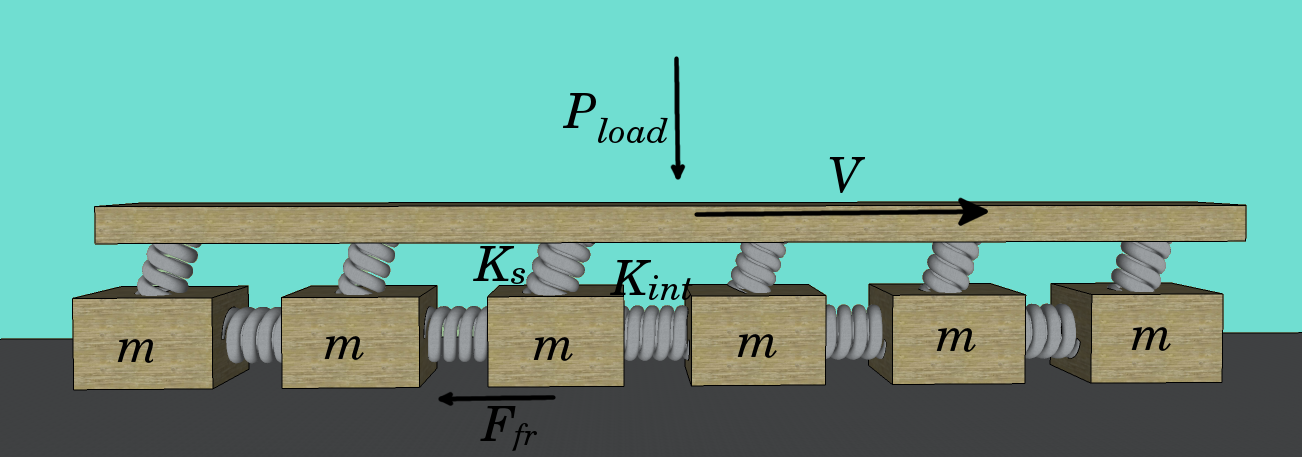}
\caption{\footnotesize Schematic of the spring-block model with the notation used in the text.}
\label{figmodel} 
\end{center}
\end{figure}

As stated in the introduction, the purpose of this work is to investigate the variation of 
friction coefficients in the presence of structured surfaces, also taking into account material 
elasticity. With this in mind, we start from a one dimensional spring-block model. 
This model was first introduced in 1967 by Burridge and Knopoff \cite{burr} in the study of the 
elastic deformation of tectonic plates. Despite its simplicity, the model is still used not only in 
this field \cite{equake1}-\cite{equake3}, but also to investigate some aspects of dry friction on 
elastic surfaces, e.g. the static to dynamic friction transition 
\cite{braun}-\cite{urb}, stick-slip behaviour \cite{capoz1}-\cite{sche} and the role of regular 
patterning \cite{capoz3}.

The model is illustrated in figure \ref{figmodel}: an elastic body, sliding along a rigid surface, 
is discretized in a chain of blocks of mass $m$ connected by springs of stiffness $K_{int}$, 
attached to a slider moving at constant velocity $v$ by means of springs of stiffness $K_{s}$ to 
take into account shear deformation. The surface of the sliding plane is considered as a first 
approximation homogeneous and infinitely rigid. 

Friction between the blocks and the surface can be introduced in many ways: for example in 
\cite{braun} it is modeled through springs that can attach and detach during 
motion. However, in our study we will use a classical AC friction force between blocks and 
surface through microscopic friction coefficients, as it is done, for example, in \cite{trom}. In 
this way it is possible to directly introduce a pressure load as in the figure. Hence, on 
each block the acting forces are:
\begin{enumerate}[(i)]
 \item The shear elastic force due to the slider uniform motion, $F_s = K_s (vt+l_i-x_i)$, 
where $x_i$ is the position of the block $i$ and $l_i$ is its rest position.
 \item The internal elastic restoring force between blocks $F_{int} = K_{int} 
(x_{i+1}+x_{i-1}-2x_i)$.
 \item The normal force $F_n$, which is the total normal force divided for the number of blocks in 
contact with the surface. 
 \item A viscous force $F_{damp} = -m \gamma \dot{x_i}$ to account for damping effects, with 
$\gamma$ chosen in the underdamped regime.
 \item The AC friction force $F_{fr}$: if the block $i$ is at rest, the friction force is equal 
and opposite to the resulting moving force, up to the threshold $F_{fr} = {\mu_s}_i \; {F_n}$. 
When this limit is exceeded, a constant dynamic friction force opposes the motion, i.e. $F_{fr} = 
{\mu_d}_i \; F_n$.
The microscopic friction coefficients of each block, namely ${\mu_s}_i$ and 
${\mu_d}_i$, are assigned through a Gaussian statistical dispersion to account for the random 
roughness of the surface. Thus, the probability distribution for the static coefficient is $ 
p({\mu_s}_i) = (\sqrt{2\pi}\sigma_{s})^{-1} \exp{[-({\mu_s}_i-(\mu_s)_m)^2/(2\sigma_{s}^2)]} $, 
where $(\mu_s)_m$ denotes the mean microscopic static coefficient and $\sigma_s$ is its standard 
deviation. The same distribution is adopted for the dynamic coefficient ( substituting subscript 
$d$ to $s$). The macroscopic friction coefficients, obtained through the sum of 
all the friction forces on the blocks, will be denoted as $(\mu_s)_M$ and $(\mu_d)_M$.

\end{enumerate}
Hence, we have a system of equations for the block motion that can be solved numerically with 
a fourth-order Runge-Kutta algorithm. Since the friction coefficients of the blocks 
are randomly extracted at each run, the final result of any observable consists on an 
average of various repetitions of the simulation. Usually, we assume an elementary integration time 
step of $h=10^{-4}$ ms and we repeat the simulation about twenty times.

In order to relate the model to a realistic situation, we fix the macroscopic quantities, i.e. the 
global shear modulus $G=5$ MPa, the Young's modulus $E=15$ MPa, the mass density $\rho=1.2$ 
g/cm$^3$ (typical values for a rubber-like material with Poisson ratio $\nu=0.5$), the total length 
$L_x$, the transversal dimensions of the blocks $l_y$, $l_z$ and the number of blocks $N$. These 
quantities are then related to the stiffnesses $K_{int} = E\cdot (N-1) l_{z} l_{y}/L_{x} $ and $K_s 
= G\cdot l_{y}L_{x}/ (l_{z} N)$, the length of the blocks $l_x=L_x/N$, and their mass $m=\rho 
l_xl_yl_z$. The default values of the parameters are specified in table \ref{par_table}. An example 
of the simulated friction force time evolution of the system with these values is 
shown in figure \ref{fig1}. 

\vspace{20pt}
\begin{table}[h]
\begin{center}
\begin{tabular}{lr}
\hline
parameter & default value \\
\hline
shear modulus G                   &  5 MPa  \\
elastic modulus E                 &  15 MPa \\
density $\rho$                    &  1.2 g/cm$^3$ \\
total load pressure $P_{load}$    &  1 MPa \\
damping $\gamma$                  &  10 ms$^{-1}$ \\
slider velocity $v$               &  0.05 cm/s \\
length $l_y$                      &  1 cm \\
length $l_z$                      &  0.1 cm \\
micro. static coeff. $(\mu_s)_m$  &  1.0 (1)  \\
micro. dynamic coeff. $(\mu_d)_m$ &  0.50 (1) \\
\hline
\end{tabular}
\caption{\footnotesize Values of the default parameters of the model. For the microscopic friction 
coefficients we denote in round brackets the standard deviation of their Gaussian dispersion. The 
total length $L_x$ and the number $N$ of blocks will be specified for each considered case.}
\label{par_table}
\end{center}
\end{table}

\subsection{Smooth surfaces}\label{sec_smooth}

\begin{figure}[h]
\begin{center}
\includegraphics[scale=0.4]{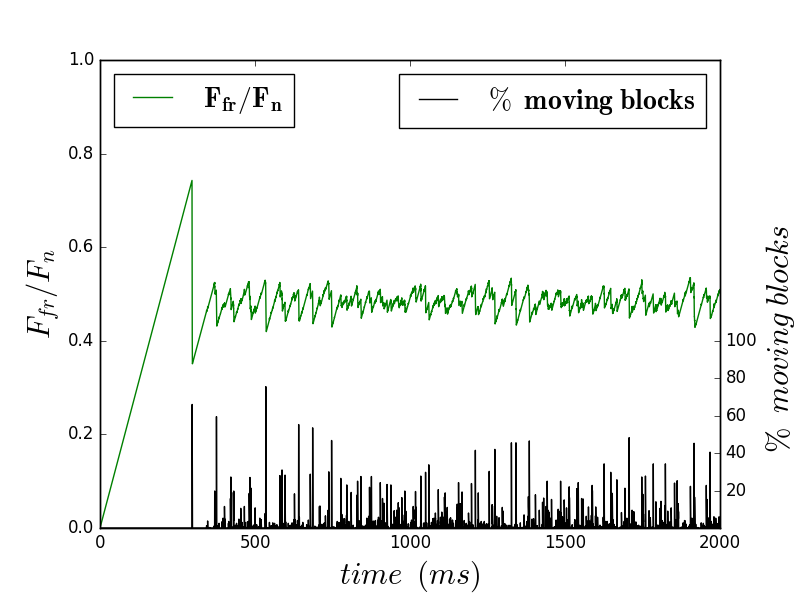}
\caption{\footnotesize Plot of the total friction force normalized with the total normal load as a 
function of time with the default set of parameters and $L_x=4.0$ cm and $N=200$. We can observe 
the typical AC force behaviour with a linear phase up to the detachment threshold followed 
by a dynamic phase with an irregular stick-slip behaviour due to the randomness of the microscopic 
friction coefficients. From this plot we can extract for example the static friction coefficient 
from the first load peak and the dynamic one from the average over the dynamic phase. The plot also 
shows the variation in time of the number of detached blocks.}
\label{fig1}
\end{center}

\begin{center}
\includegraphics[scale=0.4]{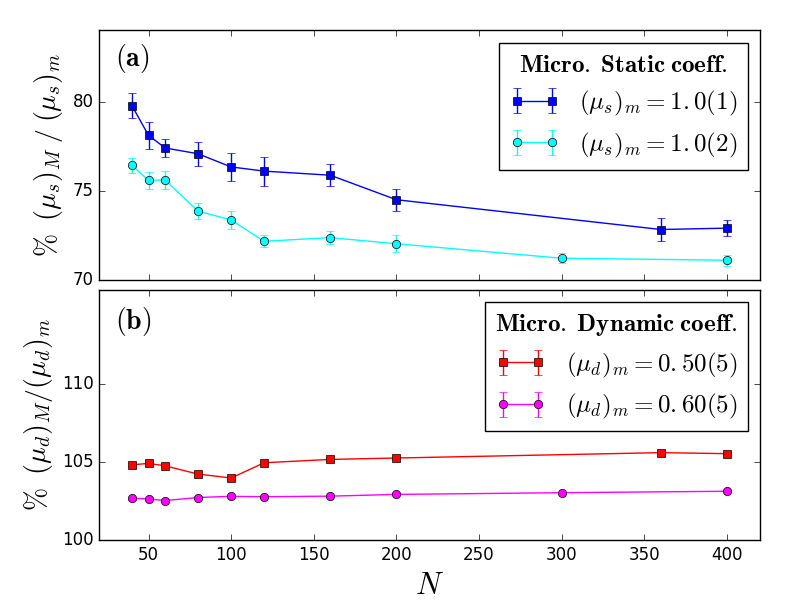}
\caption{\footnotesize Macroscopic static (a) and dynamic (b) friction coefficients using the 
default set of model parameters as a function of material discretization $N$. The block length is 
fixed at $l_x=L_x/N = 0.02$ cm, so that the ratio $K_{int}/K_s$ is also fixed. The dynamic 
coefficient is practically constant while the static coefficient slightly decreases. Two sets of 
values for the local coefficients are considered, as indicated in the legend, with the standard 
deviation of their Gaussian dispersion reported in round brackets. A wider statistical 
dispersion reduces the global static friction coefficient. 
}
\label{fig_s1}
\end{center}
\end{figure}

Before introducing surface patterning, as a preliminary study we show some results 
with the system in the standard situation of all blocks in contact. First, we show how 
the  macroscopic friction coefficients depend on microscopic ones and 
longitudinal dimensions. As seen in figure \ref{fig_s1}, with $l_x$ fixed, the 
friction coefficients decrease with the number of blocks $N$ and, consequently, the overall length 
$L_x = N l_x$. This effect is analogous to that seen in fracture mechanics, in which the global 
strength decreases with increasing element size, due to the increased statistics 
\cite{fbm1}\cite{fbm2}. Indeed, a reduction in width of the distribution of the microscopic $\mu_s$ 
leads, as expected, to an increase in the global static friction coefficient. This statistical 
argument is a possible mechanism for the breakdown of AC law,  observed for example in 
\cite{exp1}. 

The macroscopic dynamic coefficient, instead, is largely unaffected by the number of blocks, as 
shown in figure \ref{fig_s1}, and in any case its variation is less than $10\%$. We observe 
also that it is greater than the average microscopic coefficient. This is to be expected, since 
during the motion some blocks are at rest and, hence, the total friction force in the sliding 
phase has also some contributions from the static friction force (see also section \ref{an_calc}). 

On the other hand, by varying $l_x$ with fixed $N$, the values of the stiffnesses $K_{s}$ 
and $K_{int}$ are changed and, hence, the relative weight of the elastic forces. Depending on 
which one prevails, the system displays two different qualitative regimes: if 
$K_{int} > K_s$, the internal forces dominates so that, when a block begins to move after its 
static friction threshold has been exceeded, the rupture propagates to its neighbor and a 
macroscopic sliding event occurs shortly after. In this case, the total friction force in the 
dynamic phase exhibits an irregular stick-slip behaviour, as shown, for example, in figure 
\ref{fig1}. Instead, if $K_{int} \lesssim K_s$, the internal forces are less influential, so that 
the macroscopic rupture occurs only when the static friction threshold of a sufficient number of 
blocks has been exceeded. 

In a real material, the distance $l_x$ can be related to the characteristic length between 
asperities on the rough surface of the sliding material. Hence, the regime with a shorter $l_x$, 
implying a larger $K_{int}$, can be interpreted as a material whose asperities are close packed and 
slide together, while in the other limit they move independently. In the following, we will 
consider the regime $K_{int} > K_s$, which is more representative for the rubber-like parameters we 
have chosen with realistic length scales.

\begin{figure}[h]
\begin{center}
\includegraphics[scale=0.4]{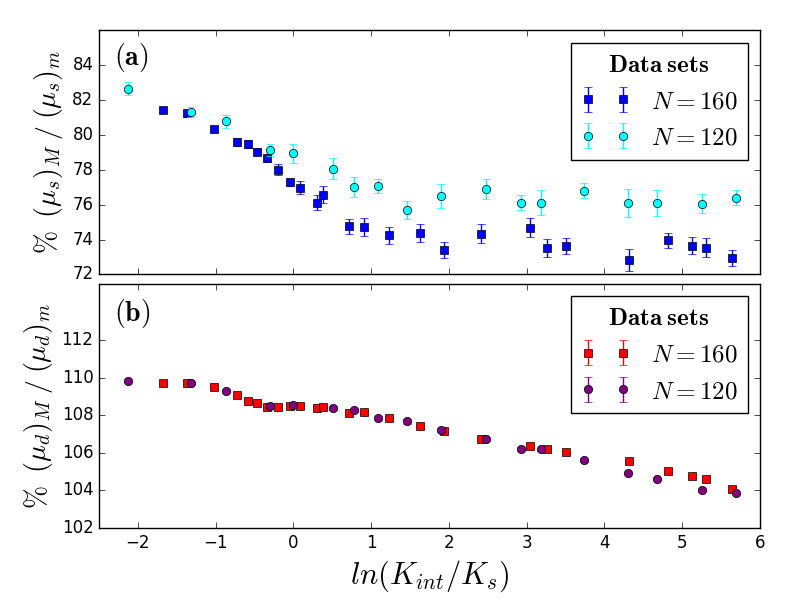}
\caption{\footnotesize Macroscopic static (a) and dynamic (b) friction coefficients using the 
default set of parameters as a function of $K_{int}/K_{s}$ obtained by varying $l_x$ for set values 
of $N$. The dynamic coefficients increase by reducing $K_{int}$. The static coefficients instead 
are constant for $K_{int} > K_{s}$, and they decrease for larger $N$, as in figure \ref{fig_s1}. 
The static coefficients begin to increase only when the stiffnesses are comparable.}
\label{fig_s2}
\end{center}
\end{figure}

The plot of the resulting macroscopic friction coefficients as a function of the stiffnesses is 
shown in figure \ref{fig_s2}: the static friction coefficients is constant in the region 
$K_{int} > K_s$ and it starts to increase when the stiffnesses become comparable. This is to be 
expected, since by reducing the force between blocks only the force due to shear 
deformation remains. The dynamic friction coefficients slightly increase by reducing 
$K_{int}$ in both the regimes for the same reason of the static coefficients: the total 
force is reduced during the sliding phase and, hence, the fraction of resting blocks is 
increased.

\subsection{Dynamic friction coefficient}\label{an_calc}

In this section, we calculate analytically the dynamic friction coefficient in the limit of 
$K_{int}=0$. This is useful as a further test of the program and to highlight some interesting 
properties of the macroscopic friction coefficient. In this limit, the blocks move independently, 
so that the resulting friction force can be obtained by averaging the behavior of a single block. 
Let us consider a single block of mass $m$, which is pulled by the slider, moving at constant 
velocity $v$, with an elastic restoring force of stiffness $k$. 
The AC friction force acts on the block, whose static and dynamic friction coefficient 
are ${\mu_s}_i$ and ${\mu_d}_i$ respectively. In the following, we will drop the index $i$ for 
simplicity.

The maximum distance $r_0$ between the block and the slider can be found by equating the 
elastic force and the static force, ${\mu_s} F_n = k r_{0}$, where $F_n$ is the normal force 
on the block. From this distance the block starts to move under the effect of the elastic force and 
dynamic friction force, therefore we can solve the motion equation for the position $x$ of 
the block: 
\begin{equation}\label{eq1}
 \ddot{x}(t) = - \omega^{2} \; ( x(t) - vt - r_{0} ) - {\mu_d} \frac{F_n}{m}
\end{equation}
where $\omega= \sqrt{k/m}$, and $r_{0} = {\mu_s} F_n /k$ is the 
initial distance between the block and the slider. By solving the differential equation with 
initial conditions $x(0)=0$, $\dot{x}(0)=0$ we obtain:
\begin{equation}\label{eq2}
 x(t) = \frac{v}{\omega}\; ( \omega t-\sin{\omega t} ) + 
\frac{F_n}{k} \; ({\mu_s}-{\mu_d}) \; (1-\cos{\omega t} )
\end{equation}
\begin{equation}\label{eq3}
\dot{x}(t) = v \; (1-\cos{\omega t} ) + \frac{\omega F_n}{k} \; (\mu_s-\mu_d) \; \sin{\omega t}
\end{equation}
The equation for the velocity (\ref{eq3}) can be used to find the time duration $T_d$ of the 
dynamic phase. After this time the block halts and the static friction phase begins. Hence, 
by setting $\dot{x}(t) =0$ and solving for $t>0$, after some manipulations using 
trigonometric relations, we find:
\begin{equation}\label{eq4}
T_d = \frac{2}{\omega} \Bigg[ \pi - \arctan{\left( (\mu_s-\mu_d)\frac{\omega F_n}{k v}\right)} 
\Bigg]
\end{equation}
In order to characterize the static phase, it is important to calculate the distance $\Delta 
r$ between the slider and the block at the time $T_d$, because the subsequent duration $T_s$ of 
the static phase will be determined by the time necessary for the slider to again reach the maximum 
distance $r_{0}$. Therefore, we must calculate $\Delta r \equiv x(T_d)-v T_d-r_{0}$. After 
some calculations we find:
\begin{equation}\label{eq5}
\Delta r = \frac{F_n}{k} \;(\mu_s-2 \mu_d )
\end{equation}
Hence, if $2 \mu_d =  \mu_s$ the block exactly reaches the slider after every dynamic phase. If $2 
\mu_d <(>) \; \mu_s$ the block stops after (before) the slider position. Hence there are two 
regimes determined by the friction coefficients. From this we can calculate the distance required by
the slider to again reach the maximum distance $r_0$ and, hence, the duration time $T_s$ 
of the static phase:
\begin{equation}\label{eq6}
T_s = \frac{2 F_n}{k v} \;(\mu_s-\mu_d )
\end{equation}
Now we have all the ingredients to calculate the time average of the friction force, from which the 
dynamic friction coefficient can be deduced. We restore the index $i$ to distinguish the 
microscopic friction coefficients from the macroscopic one $(\mu_d)_M$. We write 
the time average of the friction force as the sum of the two contributions from the dynamic phase 
and the static one:
\begin{equation}
(\mu_d)_M \equiv \frac{< F_{fr} >}{F_n} = \frac{1}{F_n} \left( \frac{T_d}{T_{tot}} {\mu_d}_i F_n + 
\frac{T_s}{T_{tot}} < F_{stat} > \right)
\end{equation}
so that:
\begin{equation}\label{eq7}
(\mu_d)_M  =  {\mu_d}_i  + \frac{T_s}{T_{tot}} \left( \frac{ < F_{stat} >}{F_n} - {\mu_d}_i \right)
\end{equation}
where $T_{tot}=T_d+T_s$ and $F_{stat}$ is the static friction force. In the static phase the 
friction force is equal to the elastic force, hence, in practice, we must calculate the time 
average of the modulus of the distance between block and slider in the static phase, i.e. the 
average of $k|vt + \Delta r| $ over the time necessary for the slider to go from $\Delta r$ to 
$r_0$. For this 
reason, we must distinguish the two regimes depending on the sign of $\Delta r$ calculated in 
equation (\ref{eq5}). After some calculations we find:
\begin{align}\label{eq8}
< F_{stat} > =  \left\{ \begin{array}{ll} 
F_n \; {\mu_d}_i \;\; \mbox{if}  \;\; 2 {\mu_d}_i \geq {\mu_s}_i \\
 \\
F_n \frac{({\mu_s}_i - 2{\mu_d}_i)^{2} + {{\mu_s}_i}^2}{4 ({\mu_s}_i - {\mu_d}_i)} \;\; \mbox{if} 
\;\;2 {\mu_d}_i < {\mu_s}_i 
\end{array} \right.
\end{align}
Finally, by substituting equation (\ref{eq8}) into (\ref{eq7}), we obtain:
\begin{align}\label{eq9}
(\mu_d)_M =  \left\{ \begin{array}{ll} 
 {\mu_d}_i \;\; \mbox{if}  \;\;\; 2 {\mu_d}_i \geq {\mu_s}_i \\
 \\
 {\mu_d}_i + \frac{T_s}{T_{tot}} \frac{( {\mu_s}_i-2{\mu_d}_i )^2}{2({\mu_s}_i-{\mu_d}_i )}\;\; 
\mbox{if} \;\;2 {\mu_d}_i < {\mu_s}_i 
\end{array} \right.
\end{align}
showing that the limit case ${\mu_d}_i = {\mu_s}_i/2$ of the two expressions coincides. Now, 
if we have $N$ non-interacting blocks, we can average the equations (\ref{eq9}) over the index $i$ 
in order to calculate the macroscopic friction coefficient in terms of the mean microscopic ones, 
$(\mu_s)_m$ and $(\mu_d)_m$. Since the second term of equation (\ref{eq9}) contains a 
complicated expression, this can be done exactly only numerically. Nevertheless, we can 
deduce that, at least in the regime of negligible $K_{int}$, (i) the resulting dynamic friction 
coefficient is always greater or equal to the microscopic one, and (ii) there are two regimes 
discriminated by the condition $2 \;(\mu_d)_m \gtrless (\mu_s)_m$. We observe also 
that in the case $2\;(\mu_d)_m \simeq (\mu_s)_m$, owing to the statistical dispersion 
of the coefficients, each block could be in both the regimes, so that the final result will be 
an average between the two conditions of (\ref{eq9}).

\begin{figure}[h]
\begin{center}
\includegraphics[scale=0.4]{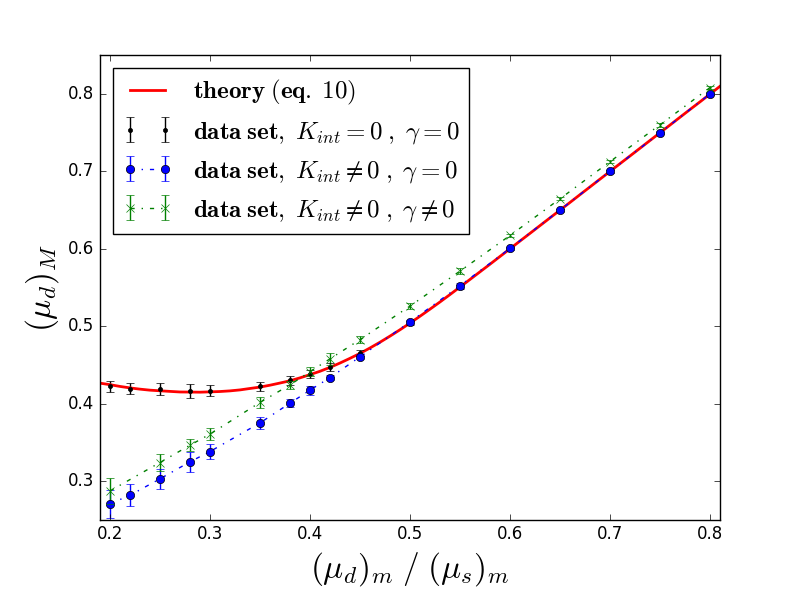}
\caption{\footnotesize Macroscopic dynamic friction coefficient as a function of the microscopic 
one. The red curve shows the theory prediction by Eq. (\ref{eq9}). Results of the ideal case 
$K_{int}=0$ are the black dots, which follow exactly the predictions. The yellow and the blue 
dots show the data sets with blocks interaction, with and without the damping $\gamma$, 
respectively. The data are obtained with the default system parameters, $N=200$ and $L_x=4.0$ cm.
}
\label{fig2}
\end{center}
\end{figure}

The following plot (figure \ref{fig2}) shows the behavior predicted by equation (\ref{eq9}) 
compared with the simulations either in ideal case $K_{int}=0$, that perfectly match with the 
theory, and with blocks interactions, that diverge from the predictions only for 
$(\mu_d)_m/(\mu_s)_m<0.5$. This is to be expected, since the internal forces become much more 
influential if the blocks can move without a strong kinetic friction.

\section{Structured Surfaces} \label{sec_patt}

\begin{figure}[h]
\begin{center}
\includegraphics[scale=0.20]{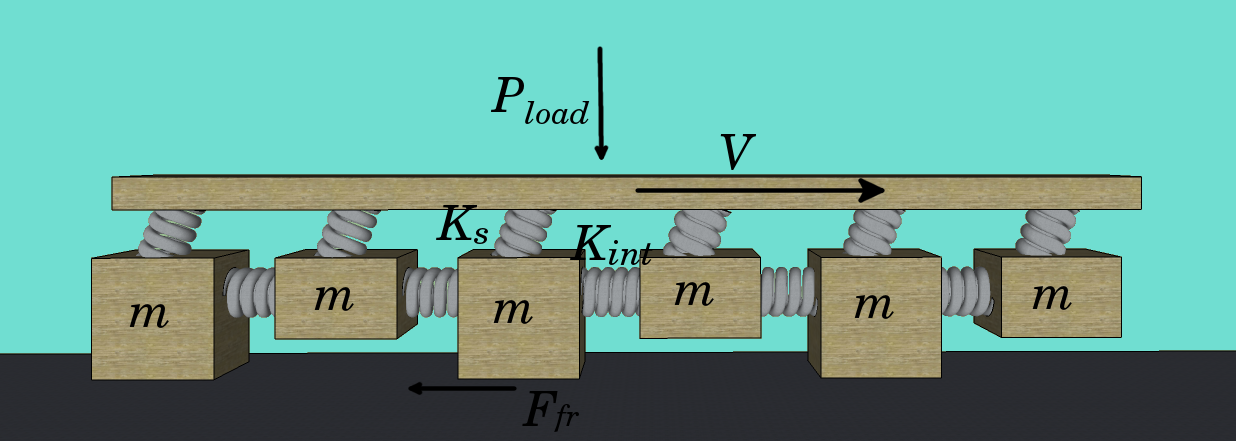}
\caption{\footnotesize In order to evaluate the effect of surface structuring, we assume in the 
spring-block model that a number of blocks is no longer in contact with the sliding plane and is 
instead free to oscillate. The total pressure is maintained constant, so that the normal force on 
the blocks in contact increases.}
\label{figmodel2}
\end{center}
\end{figure}

\subsection{First level patterning}\label{sec_patt1}

Next, we set to zero the friction coefficients relative to some blocks in order to simulate the 
presence of structured surfaces on the sliding material (figure \ref{figmodel2}). We start with a 
periodic regular succession of grooves and pawls. This pattern has already been 
studied both experimentally \cite{brass}\cite{snake} and numerically with a slightly different 
model \cite{capoz3}. Our aim is therefore to first obtain known results so as validate the model.

We consider a succession of $N_g$ grooves of size $L_g$ at regular distances of $L_g$, so that 
only half of the surface is in contact with respect to previous simulations. The number of blocks 
in each groove is $n_g = N/(2 N_g)$, and $L_g=n_g L_x/N$. The friction coefficients of these blocks 
are set to zero while default values are used for the remaining ones.     

\begin{figure}[h]
\begin{center}
\includegraphics[scale=0.4]{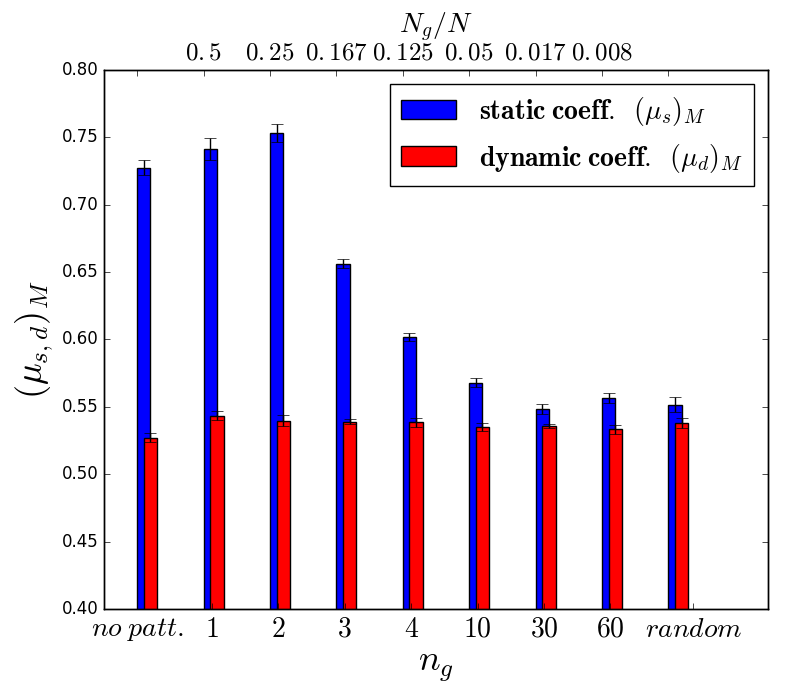}
\caption{\footnotesize Static and dynamic friction coefficients for a periodic regular 
patterned surface as a function of number of blocks in a groove $n_g$. Results are obtained 
with the default set of parameters, $L_x=7.2$ cm and $N=360$, and microscopic friction coefficients 
$(\mu_s)_m=1.0(1)$ and $(\mu_d)_m=0.50(1)$. The behaviour is analogous to that observed in the 
literature \cite{snake}. }
\label{fig_p1}
\end{center}
\end{figure}

Figure \ref{fig_p1} shows that, as expected, the static friction coefficient decreases with larger 
grooves while the dynamic coefficient is approximately constant. In the case of small grooves, e.g. 
for $n_g \leq 2$, there is no reduction, confirming the results in \cite{capoz3}, where it is found 
that the static friction reduction is expected only when the grooves length is inferior to a 
critical length depending on the stiffnesses of the model. The critical length can be rewritten 
in terms of the adimensional ratio $N_g/N$ and, by translating it in the context of our model, we 
obtain $(N_g/N)_{cr} = 2 \sqrt{K_{s}/K_{int}}$. For the data set of figure \ref{fig_p1} the 
critical value is $(N_g/N)_{cr} \simeq 0.23$ and, indeed, our results display static friction 
reduction for groove size whose $N_g/N$ is inferior to this. 

\begin{figure}[h]
\begin{center}
\includegraphics[scale=0.3]{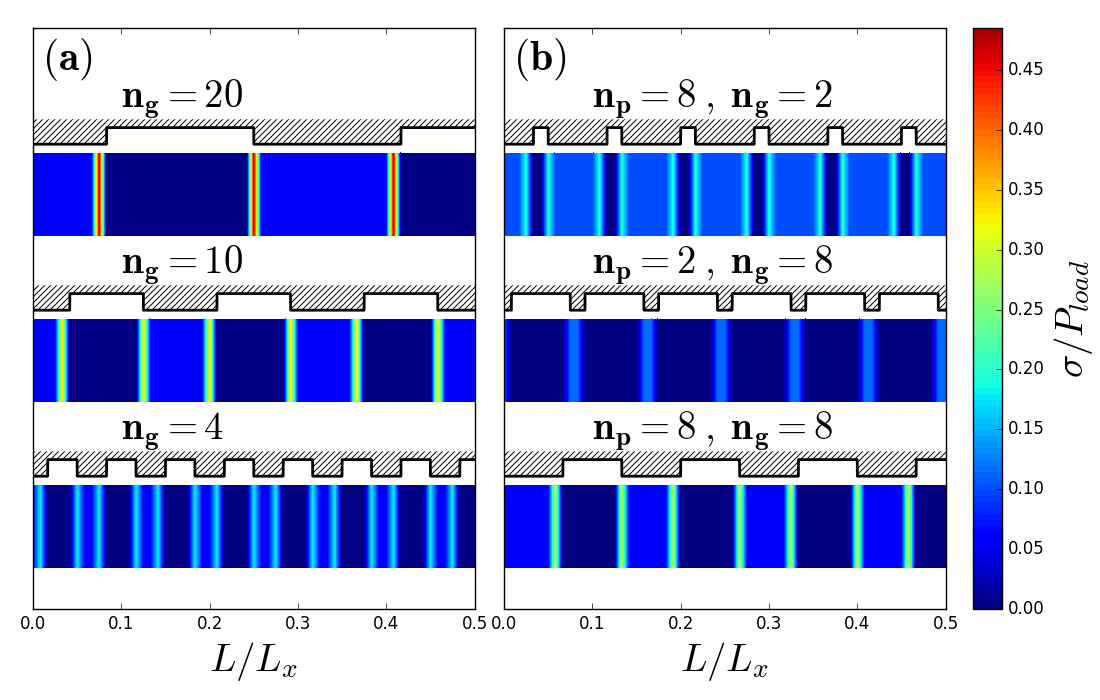}
\caption{\footnotesize Normalized stress distribution $\sigma$ as a function of the longitudinal 
distance $L$ along the patterned surface, i.e. stress acting on each block normalized by the total 
applied pressure. For the three considered cases, the patterning profile is illustrated with a 
black line.  
\newline
a) Case of regular periodic patterning for different $n_g$ values. For larger grooves, the stress 
on the blocks at the pawl edges increases. 
\newline 
b) Three cases varying the relative length of grooves and pawls: for small pawls, despite the 
larger grooves, the stress is reduced because of the increased normal load on the blocks in contact.
}
\label{fig3}
\end{center}
\end{figure}

The origin of this behaviour in the spring-block model can easily be understood by looking at the 
stress distribution on the patterned surfaces (figure \ref{fig3} a): stresses at the edge of the 
pawls increase with larger grooves, so that the detachment threshold is exceeded earlier and the 
sliding rupture propagates starting from the edge of the pawls. The larger the grooves, the more 
stress is accumulated. Thus, for a constant number of blocks in contact, i.e. constant real contact 
area, the static friction coefficient decreases with larger grooves.

\begin{figure}[h]
\begin{center}
\includegraphics[scale=0.4]{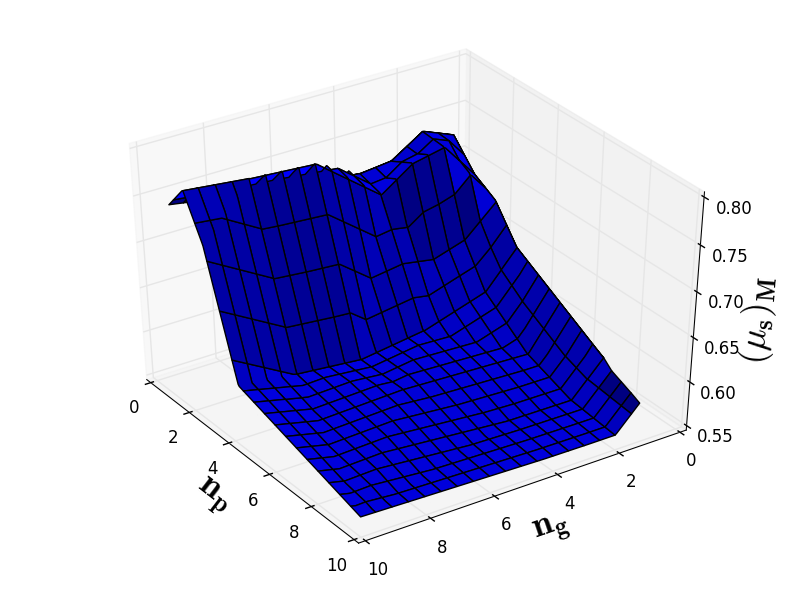}
\caption{\footnotesize Plot of the static friction coefficient as a function of the size of 
grooves and pawls (see table \ref{tpat0}). }
\label{fig4}
\end{center}
\end{figure}

Next, we evaluate configurations in which the pawls and the grooves have different sizes, i.e. 
the fraction of surface in contact is varied. This is equivalent to changing the normal force 
applied to the blocks in contact, since the total normal force is fixed. We must denote these 
single-level configurations with two symbols, the number of blocks in the grooves $n_g$ as 
previously, but also the number of blocks in the pawls, namely $n_p$. When they are the same we 
will report only to $n_g$, as in figure \ref{fig_p1}. All the results obtained for the macroscopic 
friction coefficient are reported in table \ref{tpat0} and shown in figure \ref{fig4}.

We can observe that, for a given fraction of surface in contact, the static friction 
decreases for larger grooves, as in the case ${n_p}={n_g}$. However, with different ${n_p}$ and 
${n_g}$ values, static friction increases when the pawls are narrower than the grooves, i.e. when 
the real contact area is smaller than one half. 

This would appear to be in contrast with results observed previously relative to the relative 
groove size. However, the normal load applied to the blocks must also be taken into account: if the 
normal force is distributed on fewer blocks, the static friction threshold will also be greater 
although the driving force is increased (figure \ref{fig3} b). Hence, static friction reduction 
due to larger grooves can be balanced by reducing the real contact area, as highlighted by results 
in table \ref{tpat0}.

The interplay between these two concurrent mechanisms explains the observed behaviour of the 
spring-block model with single-level patterning using pawls and grooves of arbitrary size.

The dynamic friction coefficient, on the other hand, displays reduced variation in the presence of 
patterning, and in practice increases only when there is a large reduction of the number of blocks 
in contact.

Finally, we have also tested a configuration with randomly distributed grooves, i.e. half the 
friction coefficients of randomly chosen blocks are set to zero. This turns out to be the 
configuration with the smallest static friction. This can be explained by the fact that there are 
grooves and pawls at different length scales, so that it is easier to trigger sequences of 
ruptures, leading to a global weakening of the static friction. Thus, the simulations show that in 
general a large statistical dispersion in the patterning organization is detrimental to the static 
friction of a system, whilst an ordered structure is preferable in most cases.
\vspace{20pt}

\subsection{Hierarchical patterning} \label{sec_patt2}
 
\begin{figure}[h]
\begin{center}
\includegraphics[scale=0.15]{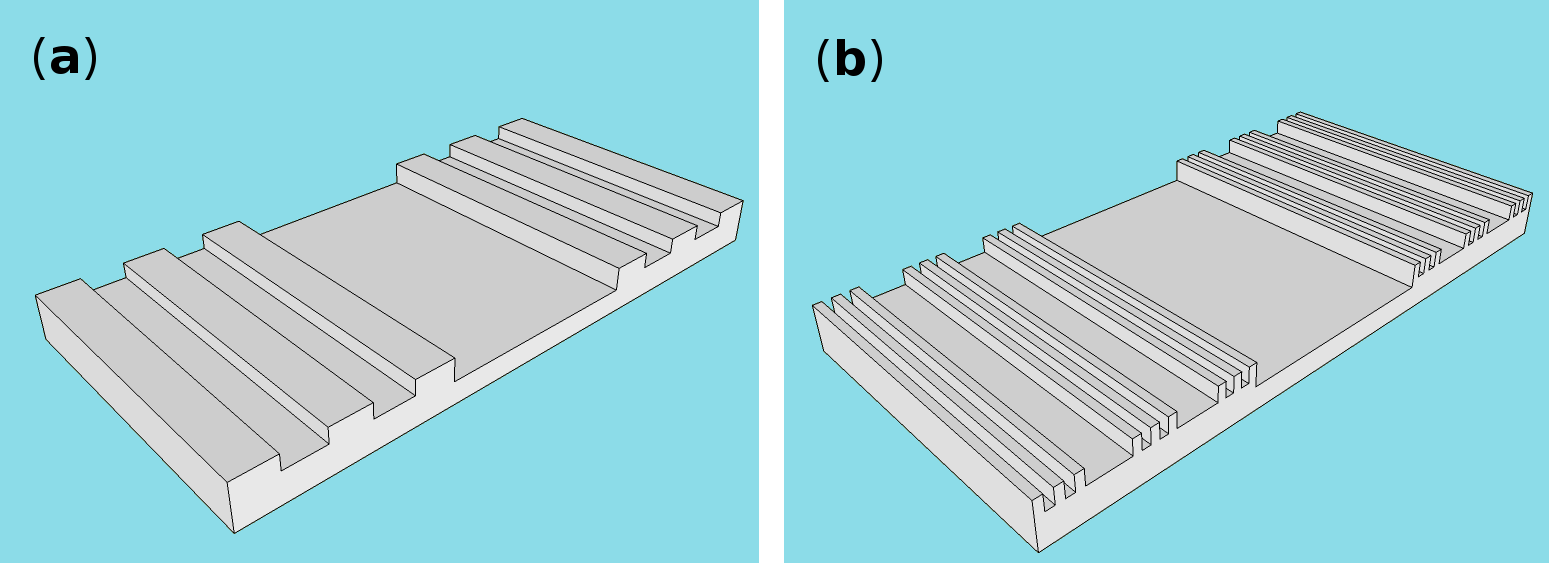}
\caption{\footnotesize Example of the elementary structure of surfaces with two (a) and 
three (b) levels of patterning. With the notation used in the text the left configuration 
is denoted by $\lambda_1=1/3$, $\lambda_2=1/15$, the right one by $\lambda_1=1/3$, 
$\lambda_2=1/15$, $\lambda_3=1/75$.}
\label{figmodel3}
\end{center}
\end{figure}

We now consider grooves on different size scales, arranged in 2- and 3-level hierarchical 
structures, as shown in figure \ref{figmodel3}. Further configurations can be constructed with more 
hierarchical levels, by adding additional patterning size scales.

\begin{figure}[h!]
\begin{center}
\includegraphics[scale=0.4]{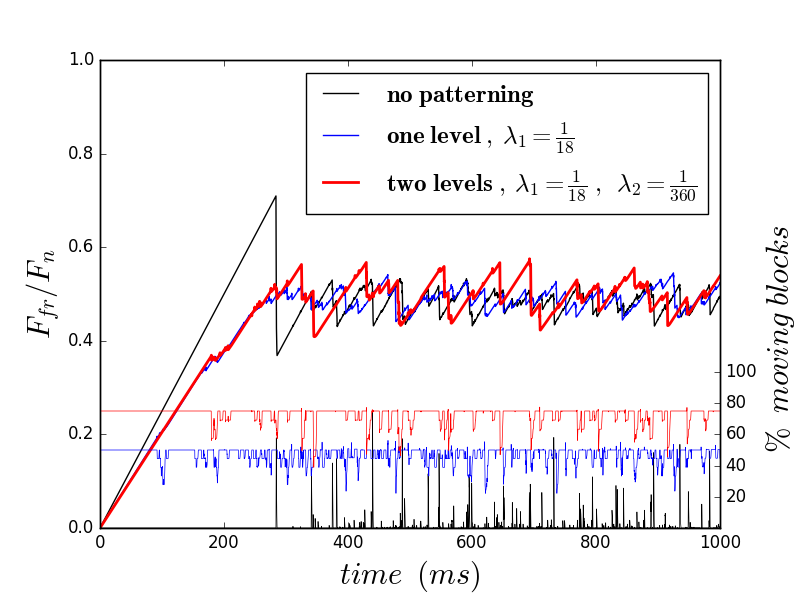}
\caption{\footnotesize Comparison of the total friction force normalized with the total load 
for increasing levels of patterning, using the default set of parameters and $L_x=7.2$ cm and 
$N=360$. A reduction of the static friction force is observed with respect to the non patterned 
case. However, adding a further level, the static friction increases again, and dynamic friction 
displays a more evident time variation although the average is approximately the same.
}
\label{fig5}
\end{center}
\end{figure}

The configurations are identified using the ratios between the length of the grooves at 
level $i$ and the total length: $\lambda_i \equiv L_{g}^{(i)}/L_x$. For example, a hierarchical 
configuration indicated with $\lambda_1=1/5$, $\lambda_2=1/15$, $\lambda_3=1/120$ has three levels 
with groove sizes $L_{g}^{(1)} = L_x/5$, $L_{g}^{(2)} = L_x/15$, $L_{g}^{(3)} = L_x/120$, 
respectively, from the largest to the smallest. In the spring block model this implies that the 
number of blocks in  each groove at level $i$ is $n_{g}^{(i)} = N \lambda_i$. For readability, 
these numbers will be shown in the tables. Macroscopic friction coefficients for various 
multi-level configurations are reported in the appendix \ref{appendix}. The comparison of the total 
friction force as a function of the time between the case of a smooth surface, single level and 
two-levels patterning is shown in figure \ref{fig5}.

\begin{figure}[h!]
\begin{center}
\includegraphics[scale=0.4]{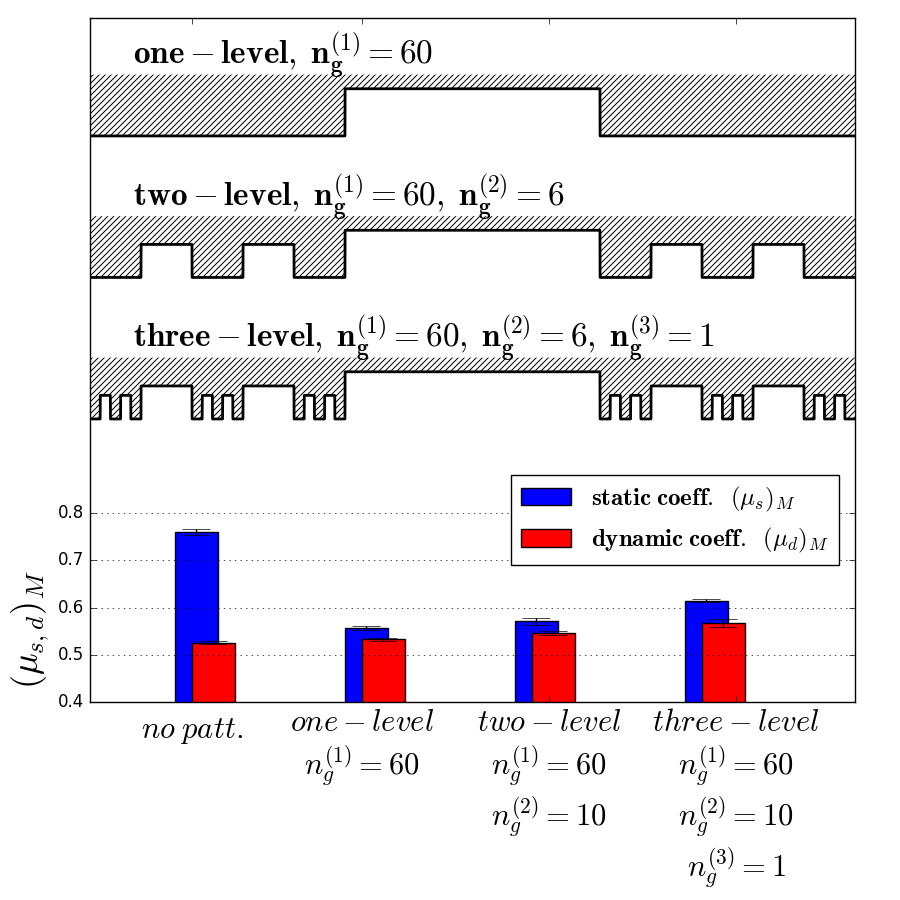}
\caption{\footnotesize Static and dynamic friction coefficients obtained by adding 
further hierarchical levels (labeled as in table \ref{tpat1}) but keeping fixed the first level 
size. For each considered case the grooves profile along the surface is shown. Results are 
obtained with the default set of parameters, $L_x=2.4$ cm, $N=120$ and microscopic friction 
coefficients $(\mu_s)_m=1.0(1)$ and $(\mu_d)_m=0.50(1)$. 
}
\label{fig_p2}
\end{center}
\end{figure}

\begin{figure}[h!]
\begin{center}
\includegraphics[scale=0.4]{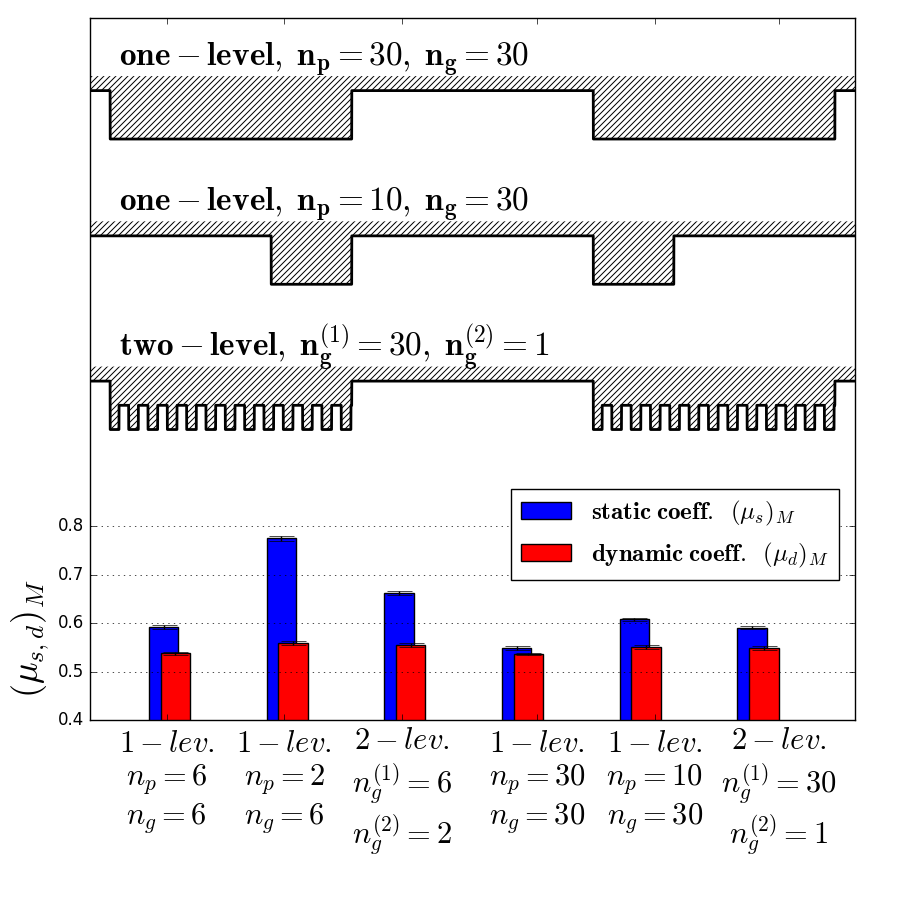}
\caption{\footnotesize Comparison of static and dynamic friction coefficients for some cases of 
one- and two-level patterning with the default parameters, $L_x=7.2$ cm, $N=360$. For each 
considered case the groove profile along the surface is shown. With two levels 
the static friction is reduced with respect to the corresponding one-level case with same contact 
area and same size of the largest grooves. 
}
\label{fig_p3}
\end{center}
\end{figure}

In general, by adding more levels of patterning (as in figure \ref{figmodel3}) the static friction 
coefficients increases with respect to the single-level configuration whose groove size is that of 
the first hierarchical level (see figure \ref{fig_p2}). This effect is due to the increased normal 
force on the remaining contact points, since the total normal force applied to the whole surface is 
constant, but it is distributed on a smaller number of blocks. This increase of the 
static friction becomes more significant the more the length scales of the levels are different. 
Indeed, if the groove size of the first level is fixed, there is a progressive reduction of the 
static friction as the second-level groove size increases, down to the value obtained with a single 
level. These trends can be clearly observed in tables \ref{tpat1} and \ref{tpat2}. On the other 
hand, if we compare a hierarchical configuration with a single-level one with the same contact area, 
i.e. if we compare the results of table \ref{tpat0} and table \ref{tpat2} for the same fraction 
of surface in contact and the same first level size, we observe a reduction of the static friction 
(see figure \ref{fig_p3}). 

Hence, a multi-structured surface produces an increase in static friction with respect to a 
single-level patterning with the same first level groove size, but a decrease with respect to 
that with the same real contact area. The explanation is the following: if the normal load is 
fixed, a structure of nested grooves allows to distribute the longitudinal forces on more points of 
contact on the surface, so that the static threshold will be exceeded earlier with respect an 
equal number of points arranged without a hierarchical structure. In other words, the 
hierarchical structure increases the number of points subjected to stress concentrations at the 
edges of the grooves.

Hence the role of the hierarchy can be twofold: if the length scale of the grooves at some level is 
fixed, by adding a further hierarchical level with a smaller length scale we can strengthen the 
static friction by reducing the number of contact points. On the other hand, among the 
configurations with the same fixed fraction of surface in contact, the hierarchical one has the 
weakest static friction, because the longitudinal stress is distributed on more points.

Moreover, the dynamic friction coefficients do not show variations greater 
than a few percent with respect the case of smooth surfaces, but they increase by 
reducing the blocks in contact and, consequently, also by adding hierarchical levels.

From all of these considerations, we can also deduce that, by increasing the number of 
hierarchical levels and by appropriately choosing the groove size at each level, it is possible 
to fine tune the friction properties of a surface, exploiting an optimal compromise between the 
extremal effects. Hierarchical structure is essential as it provides the different length scales 
needed to manipulate the friction properties of the surface.

\section{Conclusions}\label{conc}

In this paper, we have investigated how the macroscopic friction coefficients of an elastic 
material are affected by a multilevel structured surface constructed with patterning at different 
length scales. Our results were obtained by means of numerical simulations using a one-dimensional 
spring-block model, in which friction is modeled using the classical AC friction force 
with microscopic friction coefficients assigned with a Gaussian statistical distribution. System 
parameters were chosen in such a way as to be as close as possible to realistic situations for 
rubber-like materials sliding on a rigid homogeneous plane.

Tests were initially performed for a smooth surface: in this case the model predicts that 
the friction coefficients slightly decrease with the number of asperities in contact, i.e. the 
number of blocks, an effect that can be ascribed to statistical dispersion. The friction 
coefficients also decrease if the asperities are close-packed (i.e. the blocks distance is 
shorter), because their slipping occurs in groups and the local stress is increased .

The presence of patterning was then simulated by removing contacts (i.e. friction) at 
selected locations, varying the length of the resulting pawls and grooves and the number of 
hierarchical levels. The model predicts the expected behaviour for a periodic regular 
patterning, correctly reproducing results from experimental studies.

We have shown that in order to understand the static friction behaviour of the system in presence 
of patterning two factors must be taken into account: the length  of the grooves and the 
discretization of the contacts (i.e. number of blocks in contact). The longitudinal force acting on 
the pawls increases with larger grooves, so that a smaller global static coefficient can be 
expected. However, if the fraction of surface in contact is small, the friction threshold increases 
and so does the global static friction. Single-level patterning frictional behaviour 
can be understood in terms of these mechanisms.

In a multi-level patterned structure, the hierarchy of different length scales provides a solution 
to reduce both the effects. If at any level of the hierarchy the grooves are so large that the 
static friction is severely reduced, a further patterning level, whose typical length scales are 
definitely smaller, can enhance it again, since with a reduced number of contact points the 
static threshold is increased. 

On the other hand, if we compare the configurations with the same fraction of surface in contact, 
the hierarchical structure has the weakest static friction, since it increases the number of points 
at the edges between pawls and grooves and, consequently, the fraction of surface effectively 
subjected to longitudinal stress concentrations. Thus, a hierarchical structure can be used to 
construct a surface with a small number of contact points but with reduced static friction.

These results indicate that exploiting hierarchical structure, global friction properties 
of a surface can tuned arbitrarily acting only on the geometry, without changing 
microscopic friction coefficients. To achieve this, it is essential to provide structuring at 
various different length scales.
 
In this study, the effect of different length scales and structure has been studied for constant
material stiffnesses and local friction coefficients, but in future we aim to verify the existence 
of universal scaling relations by changing the system size parameters, and to analyze the role of 
the mechanical properties, e.g. for composite materials or graded frictional surfaces. Also, a 
natural extension of this study is to consider two- or three-dimensionally patterned surfaces, 
allowing a more realistic description of experimental situations and a larger variety of surface 
texturing possibilities. These issues will be explored in future works.

\section*{Acknowledgments}
N.M.P is supported by the European Research Council (ERC StG Ideas 2011 BIHSNAM no. 279985 and ERC 
PoC 2015 SILKENE no. 693670), and by the European Commission under the Graphene Flagship (WP14 
'Polymer nanocomposites', no. 696656). G.C. and F.B. are supported by BIHSNAM.

\clearpage

\appendix

\section{Tables}\label{appendix}

\begin{table} [h]
\begin{center}
\begin{tabular}{|l|l|c||c|c|}
\hline
${n_p}$ & ${n_g}$  & $S/S_{tot}$ & $(\mu_s)_{M}$ & $(\mu_d)_{M}$\\
\hline
\multicolumn{2}{|l|}{no patt.}   & 1    & 0.727 (7) & 0.527 (3) \\
\hline
1  & 1   & 1/2    & 0.741 (8) & 0.543 (3) \\
2  & 2   & 1/2    & 0.753 (6) & 0.540 (4) \\
3  & 3   & 1/2    & 0.656 (4) & 0.539 (2) \\
4  & 4   & 1/2    & 0.602 (3) & 0.539 (3) \\
10 & 10  & 1/2    & 0.568 (4) & 0.535 (3) \\
20 & 20  & 1/2    & 0.563 (4) & 0.536 (3) \\
30 & 30  & 1/2    & 0.543 (4) & 0.535 (2) \\
60 & 60  & 1/2    & 0.557 (4) & 0.533 (3) \\
\hline
1  & 2   &  1/3   & 0.755 (7)  & 0.555 (5) \\
2  & 4   &  1/3   & 0.722 (9)  & 0.554 (4) \\
4  & 8   &  1/3   & 0.599 (2)  & 0.545 (4) \\
10 & 20  &  1/3   & 0.593 (4)  & 0.545 (3) \\
20 & 40  &  1/3   & 0.566 (4)  & 0.542 (4) \\
1  & 3   &  1/4   & 0.733 (9)  & 0.563 (8) \\
2  & 6   &  1/4   & 0.775 (5)  & 0.559 (5) \\
5  & 15  &  1/4   & 0.649 (4)  & 0.554 (4) \\
10 & 30  &  1/4   & 0.607 (3)  & 0.551 (5) \\
1  & 4   &  1/5   & 0.729 (10) & 0.567 (7) \\
2  & 8   &  1/5   & 0.780 (7)  & 0.564 (4) \\
4  & 16  &  1/5   & 0.684 (6)  & 0.561 (4) \\
12 & 48  &  1/5   & 0.602 (4)  & 0.548 (5) \\
1  & 8   &  1/9   & 0.744 (13) & 0.587 (8) \\
2  & 16  &  1/9   & 0.810 (7)  & 0.585 (7) \\
\hline
2  & 1   &  2/3   & 0.713 (13) & 0.534 (3) \\
4  & 2   &  2/3   & 0.624 (9)  & 0.534 (2) \\
8  & 4   &  2/3   & 0.568 (4)  & 0.532 (2) \\
20 & 10  &  2/3   & 0.547 (3)  & 0.530 (2) \\
40 & 20  &  2/3   & 0.541 (2)  & 0.531 (2) \\
3  & 1   &  3/4   & 0.699 (4)  & 0.532 (2) \\
6  & 2   &  3/4   & 0.586 (2)  & 0.531 (2) \\
15 & 5   &  3/4   & 0.548 (2)  & 0.529 (2) \\
30 & 10  &  3/4   & 0.554 (3)  & 0.528 (2) \\
4  & 1   &  4/5   & 0.663 (3)  & 0.531 (3) \\
8  & 2   &  4/5   & 0.568 (2)  & 0.530 (1) \\
16 & 4   &  4/5   & 0.544 (2)  & 0.528 (1) \\
48 & 12  &  4/5   & 0.548 (3)  & 0.527 (2) \\
8  & 1   &  8/9   & 0.602 (2)  & 0.528 (2) \\
16 & 2   &  8/9   & 0.545 (2)  & 0.528 (2) \\
\hline 
\end{tabular}
\caption{\footnotesize Table of the macroscopic friction coefficients for one level patterning 
configuration, with the default set parameters, $N=360$ and $L_x=7.2$ cm. We denote with 
${n_p}$ and ${n_g}$ the number of blocks in each pawl and groove respectively. The 
column $S/S_{tot}$ reports the fraction of the blocks still in contact with the 
surface, i.e. the real area contact.}
\label{tpat0}
\end{center}
\end{table}

\begin{table} 
\begin{center}
\begin{tabular}{|l|l|l|c||c|c|}
\hline
$n_g^{(1)}$ & $n_g^{(2)}$ & $n_g^{(3)}$ & $S/S_{tot}$ & $(\mu_s)_{M}$ & $(\mu_d)_{M}$\\
\hline
8  &    &   & 8/15 & 0.575 (7) & 0.531 (5) \\
8  & 1  &   & 4/15 & 0.617 (8) & 0.544 (4) \\
8  & 2  &   & 4/15 & 0.597 (7) & 0.545 (6) \\
\hline
24 &    &   & 3/5  & 0.569 (5) & 0.529 (3)  \\
24 & 1  &   & 3/10 & 0.598 (9) & 0.538 (4) \\
24 & 2  &   & 3/10 & 0.594 (8) & 0.538 (4) \\
24 & 8  &   & 2/5  & 0.571 (5) & 0.536 (4) \\
24 & 8  & 1 & 1/5  & 0.628 (5) & 0.550 (9) \\
24 & 8  & 2 & 1/5  & 0.621 (5) & 0.551 (9) \\
\hline
40  &    &   & 2/3  & 0.565 (4) & 0.525 (4) \\
40  & 1  &   & 1/3  & 0.608 (4) & 0.532 (5) \\ 
40  & 2  &   & 1/3  & 0.588 (4) & 0.533 (5) \\
40  & 8  &   & 2/5  & 0.573 (4) & 0.534 (5) \\
40  & 8  & 1 & 1/5  & 0.599 (5) & 0.548 (7) \\
40  & 8  & 2 & 1/5  & 0.603 (7) & 0.547 (7) \\
\hline 
\end{tabular}
\caption{\footnotesize Table of the macroscopic friction coefficients for some cases of two 
and three levels patterning, with the default set parameters, $N=120$ and $L_x=2.4$ cm. The first 
three columns show the numbers $n_g^{(i)}$, the blocks in the grooves at level $i$, in order to 
identify the configuration. The column $S/S_{tot}$ reports the fraction of the blocks still in 
contact with the surface, i.e. the real area contact.}
\label{tpat1}
\end{center}
\end{table}

\begin{table} 
\begin{center}
\begin{tabular}{|l|l|l|c||c|c|}
\hline
$n_g^{(1)}$ & $n_g^{(2)}$ & $n_g^{(3)}$ & $S/S_{tot}$ & $(\mu_s)_{M}$ & $(\mu_d)_{M}$\\
\hline
6  &    &   & 1/2  & 0.592 (5) & 0.536 (2) \\
6  & 1  &   & 1/4  & 0.662 (4) & 0.555 (5) \\
10 &    &   & 1/2  & 0.568 (4) & 0.535 (3)  \\
10 & 1  &   & 1/4  & 0.614 (7) & 0.552 (4)  \\
10 & 2  &   & 1/4  & 0.603 (4) & 0.548 (2)  \\
20 &    &   & 1/2  & 0.563 (4) & 0.536 (3)  \\
20 & 1  &   & 1/4  & 0.574 (5) & 0.550 (4)  \\
20 & 2  &   & 1/4  & 0.573 (3) & 0.551 (2) \\
30 &    &   & 1/2  & 0.543 (4) & 0.535 (2) \\
30 & 1  &   & 1/4  & 0.591 (3) & 0.549 (4) \\
30 & 2  &   & 1/4  & 0.601 (3) & 0.550 (4) \\
\hline
60  &    &   & 1/2   & 0.557 (4)  & 0.533 (3) \\
60  & 2  &   & 1/4   & 0.581 (5)  & 0.546 (3) \\
60  & 6  &   & 1/4   & 0.610 (2)  & 0.547 (3) \\
60  & 6  & 1 & 1/8   & 0.653 (7)  & 0.568 (5) \\
60  & 10  &   & 1/4   & 0.571 (7) & 0.546 (4) \\
60  & 10  & 1 & 1/8   & 0.615 (4) & 0.567 (8)\\
\hline 
\end{tabular}
\caption{\footnotesize Macroscopic friction coefficients for some cases of two- and 
three-level patterning, using the default set parameters, $N=360$ and $L_x=7.2$ cm. The first three 
columns show the numbers $n_g^{(i)}$, i.e. the number of blocks in the grooves at level $i$, in 
order to identify the configuration. The column $S/S_{tot}$ reports the fraction of surface in 
contact.}
\label{tpat2}
\end{center}
\end{table}  

\clearpage

\end{document}